\documentstyle[11pt,a4,epsf,psfig]{article}
\newcounter{sectionc}\newcounter{subsectionc}\newcounter{subsubsectionc}

\renewcommand{\section}[1] {\vspace*{0.2cm}\addtocounter{sectionc}{1}
\setcounter{subsectionc}{0}\setcounter{subsubsectionc}{0}\noindent
        {\normalsize\bf\thesectionc. #1}\par\vspace*{0.1cm}}
\renewcommand{\subsection}[1] {\vspace*{0.2cm}\addtocounter{subsectionc}{1}
        \setcounter{subsubsectionc}{0}\noindent
        {\normalsize\it\thesectionc.\thesubsectionc. #1}\par\vspace*{0.1cm}}
\renewcommand{\subsubsection}[1]

\renewenvironment{thebibliography}[1]
        {\begin{list}{\arabic{enumi}.}
        {\usecounter{enumi}\setlength{\parsep}{0pt}
\setlength{\leftmargin .50cm}{\rightmargin .0cm}
         \setlength{\itemsep}{0pt} \settowidth
        {\labelwidth}{#1.}\sloppy}}{\end{list}}
\newcommand{\nonumsection}[1] {\vspace*{0.2cm}\noindent{\normalsize\bf #1}
        \par\vspace*{0.1cm}}

\newcommand{\beq}{\begin{eqnarray}}
\newcommand{\eeq}{\end{eqnarray}}

\newcommand{\Gev}{{\rm GeV}}
\newcommand{\Mev}{{\rm MeV}}
\newcommand{\be}{\begin{equation}}
\newcommand{\ee}{\end{equation}}

\begin{document}
\setcounter{page}{1}
{}~\vglue -3cm{\vbox{ \noindent March 3, 1999\hfil {ULG-PNT-CP-98-2}
\rightline{LPTHE-Orsay-98-66}}}\vglue 1cm

\centerline{\bf PSEUDOSCALAR VERTEX, GOLDSTONE BOSON}
\centerline{\bf AND QUARK MASSES ON THE LATTICE.}

\vskip 0.8cm
\centerline{\bf{ Jean-Ren\'e Cudell$^{a}$, Alain Le Yaouanc$^{b}$
and Carlotta Pittori$^{a}$}}
\centerline{$^{a}$ Institut de Physique,
Universit\'e de Li\`ege au Sart Tilman}
\centerline{B-4000 Li\`ege, Belgique.}
\centerline{$^b$ L.P.T.H.E.\footnote{Laboratoire associe
au CNRS-URA D00063.}, Universit\'e de Paris Sud, Centre
d'Orsay,}
\centerline{91405 Orsay, France.}
\begin{abstract} We analyse the Structure Function
collaboration data on the quark pseudoscalar vertex
and extract the Goldstone boson pole contribution, in $1/p^2$. The strength
of the pole is found to be quite large at presently accessible scales.
We draw the important consequences of
this finding for the various definitions of quark masses (short distance and
Georgi-Politzer),
and point out 
problems with the operator product expansion
and with the non-perturbative renormalisation method.
\end{abstract}

\section{Continuum model for the quark pseudoscalar vertex}
\label{sec:cont}
It is well known that the quark pseudoscalar (PS) vertex contains
a non-perturba\-ti\-ve contribution from the Goldstone boson, in the continuum.
On the lattice, the use of a non perturbative renormalisation scheme
\cite{rome}
makes this contribution manifest,
although it should
go to zero for large momentum transfers. The purpose of this letter
is to extract it from lattice simulation data, and to show
that it is not negligible for presently
accessible scales. In particular, it must be subtracted when evaluating the
short distance quark
masses from the lattice through the non perturbative method
of ref.~\cite{allton}.
This method uses the off-shell
axial Ward identity
(AWI) and renormalises the mass in the
momentum subtraction (MOM) scheme of ref.~\cite{rome},
which can afterwards be related to
the $\overline{\small\rm MS}$ scheme.
The MOM renormalisation involves 
the pseudoscalar vertex, whence the necessity of
the subtraction of the Goldstone contribution
to extract the short distance quantity\footnote{
The problem would be quite different
if other methods are used
to extract quark masses, see {\it e.g.}
ref.~\cite{alpha}.}.
For physical $u,d$ quarks, the Goldstone contribution
becomes very large, larger than the perturbative part;
this corresponds to a very large dynamical $u,d$ mass,
larger than the usual current mass at the scales accessible
to standard lattice calculations.

The expected behaviour of the pseudoscalar vertex in the continuum
has been described as follows in the 70's in the works of Lane and Pagels
\cite{lane,pagels} and others.
Near the chiral limit, the one-particle-irreducible  PS  quark vertex
$\Lambda_5$ can be described through a
perturbative contribution plus a non-perturbative Goldstone boson contribution.

The perturbative contribution is of course $1 \times \gamma_5$, with
QCD radiative corrections, which according to the renormalisation group
lead to a logarithmic behaviour $[\alpha_s(p^2)]^{4/11}$ for $N_F=0.$

As to the non-perturbative contribution, firstly,
according to PCAC, the Goldstone boson must dominate other pole contributions
in
the PS vertex near the chiral limit:
the coupling of the pion to the PS vertex indeed gives
a pole~$\sim 1/(q^2+m_\pi^2)$,
where $q$ is the momentum transferred at the  vertex;
other poles (radial excitations) contributions are suppressed in the
chiral limit. At $q=0$, in terms of the
quark mass $m$,
this gives a $1/m$ pole
contribution. We emphasize that this pole contribution
explodes at $q=0$ in the  chiral limit, {\it i.e.} it is singular on the border
of the ``physical" region of momenta. This well-known fact seems
to have been underestimated, especially in its consequence
for the renormalisation of mass on the lattice near the chiral limit
(see below, section 3).

Secondly, according to the Wilson operator product expansion (OPE),
the non-perturbative
contribution  must be power behaved, {\it i.e.} at large $p^2$ it must
drop as $1/p^2$ up to logs. This
power behavior, as shown by Lane, is dictated by one of
the pion Bethe-Salpeter
amplitude in the OPE. In the latter, the dominant operator is
the vacuum to pion matrix element of the pseudoscalar density. This
leads to $1/p^2$ by the canonical dimensions and further to logs,
 $[\alpha_s(p^2)]^{7/11}$ at $N_F=0$
\footnote{In the work of Lane, as well as in
some subsequent works, one quotes
another power of logs: $[\alpha_s(p^2)]^{-4/11}$. This latter power corresponds
to the
anomalous dimension of the local operator $\bar{\psi} \gamma_5 \psi$. But one
requires a gluon exchange at lowest order, which gives an additional $\alpha_s$
factor. This was shown by Politzer \cite{politzer} for the corresponding
$\bar{\psi} \psi$ contribution to the propagator}.

Finally, a very
important relation, emphasized by Lane and Pagels and derived
from the Ward identity, connects directly the forward pseudoscalar
vertex to the scalar part of the propagator (they are essentially
proportional)\footnote{Let us also recall that in the context of the lattice,
this relation
has been discussed and used by the Rome group to fix $Z_A$ on quark states
\cite{vladikas}.}.
Hence the study of propagator OPE by Politzer \cite{politzer}
closely parallels the above considerations on the PS vertex, with the
dominant non-perturbative (power) contribution corresponding
to the quark condensate, which corresponds to the Goldstone pole contribution
in the PS vertex. We postpone the
discussion
of the propagator lattice data, which involves delicate problems
of improvement, to another paper \cite{cudell}. Nevertheless,
in the last part of the letter, using only the vertex lattice data,
we draw
important consequences for the propagator thanks to this Ward identity.

In addition, we use the Ward identity to borrow the two-loop corrections to the
above logarithmic factors, from those calculated in the
case of the propagator by the Rome group \cite{allton},
and by Pascual and de Rafael \cite{pascual}.

In the following, we will show that the lattice data behave as expected from
the continuum theoretical expectations.
To relate lattice to physical
numbers, we will use the lattice unit $a^{-1}=1.9$~GeV at $\beta=6.0$ and
 $a\Lambda_{\small\overline{\small MS}}=0.174$ for the QCD scale parameter in
the quenched
approximation.
We must stress that nothing in our discussion
depends critically on the precise values of  these
parameters:
this work is not oriented
towards accurate numerical determinations, but rather towards
questions of principle.

\section{Fit of lattice data for $\Gamma_5$  as function of the hopping
parameter $\kappa$}
Recently \cite{rakow}, Paul Rakow presented, on the behalf of the Structure
Function (QCDSF) collaboration, the
bare vertex $\Gamma_5$
data at $\beta=6.0$ (Fig.~1 of \cite{rakow}) through the product
\beq
 am_q~\tilde\Gamma_5(p^2)\equiv C\times ~am_q ~\Gamma_5 (q=0,p^2)\label{gamma5}
\eeq
 as a function of $a^2 p^2$ at three $\kappa$ values, with:
\beq \Gamma_5\equiv {Tr(\gamma_5 \Lambda_5)\over 4}\eeq
This is the only contribution that survives at $q=0$ in the
continuum limit.
By construction, $C$ is a constant with the value \cite{rakow}:
\beq
C=0.75,
\eeq
defined in \cite{rakow}
so that the numbers in r.h.s. of eq.~(\ref{gamma5}) equate approximately the
lattice data for the scalar  part of the bare improved
propagator
(same Fig.~1 of \cite{rakow}) on some range of momenta, increasing with
$\kappa$. This approximate
equality corresponds to the Ward identity to be discussed in section 4, but we
are not concerned with it, as we rely only on the vertex data.
The QCDSF data were obtained with  the SW improved fermionic
action \cite{SW}
at $c_{SW}=1.769$, $N_F=0$,
in the Landau gauge, and
$p^2$ denotes the following lattice definition of the momentum squared,
\beq
a^2p^2\equiv  4W\equiv 4\sum_\lambda \sin^2\left({ap_\lambda\over 2}\right)
\eeq
where $a$ is the lattice unit. This notation will be used herefrom, and
identified with the continuum $p^2$. We take $a^{-1}=1.9\pm 0.1$ GeV at
$\beta=6.0$ \cite{BS}.
We also use the standard lattice definition of the bare mass:
\beq
am_q={1\over 2\kappa}-{1\over 2\kappa_c}
\label{amq}
\eeq

Let us write
\beq
 am_q \tilde\Gamma_5 (p^2)=A(p^2)+am_q B(p^2)
\label{eq:fitform}
\eeq
 We shall first perform
an extrapolation linear in $m_q$ of the three datasets
to the chiral limit at $\kappa_c=0.1352$,
for each value of $p^2$.  This fit gives us both $A(p^2)$ and
$B(p^2)$, as shown in Figs.~1 and 2.
\begin{figure}[t]
\vglue 3.9cm
\hglue 5.5cm (a)\hglue 7.0cm (b)\\
\vglue -3.4cm
\centerline{
\hbox{
\psfig{file=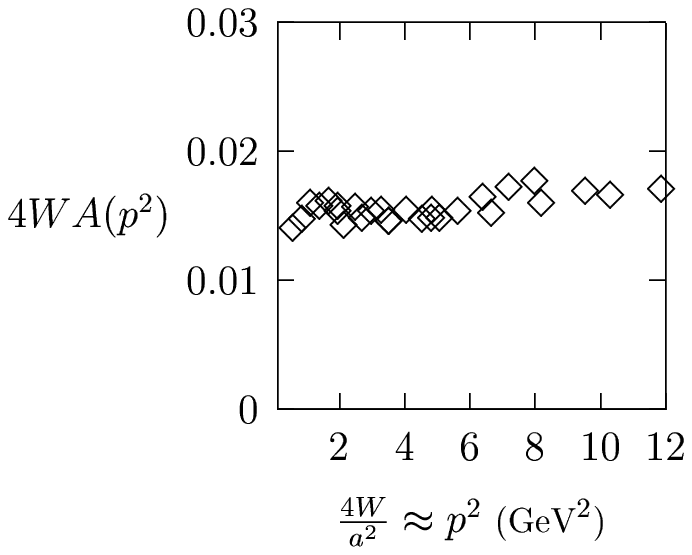,height=2.0in}
}\hglue 0.5cm \psfig{file=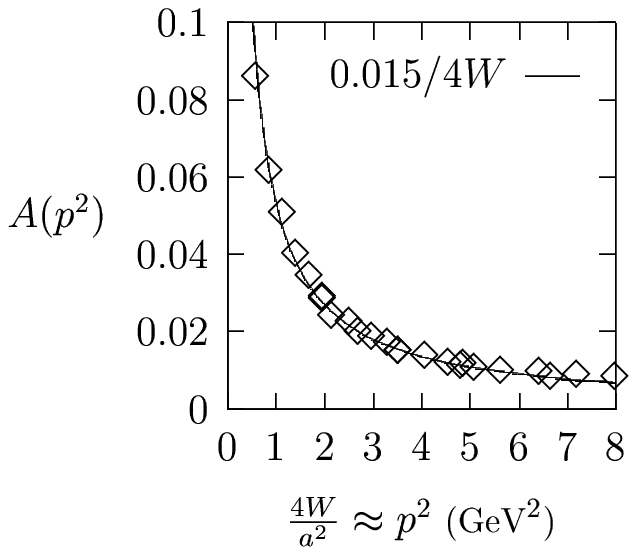,height=2.0in}
}
\caption{(a) The value of the coefficient of $a^2p^2A(p^2)$ in
eq.~(\ref{eq:fitform})
from the lattice data extrapolated at $\kappa_c$;
(b) Our fit to the extrapolated data.
}
\end{figure}
\begin{figure}[t]
\vglue 1.cm
\hglue 5.0cm (a)\hglue 6.4cm (b)\\
\vglue -1.2cm
\centerline{
\psfig{file=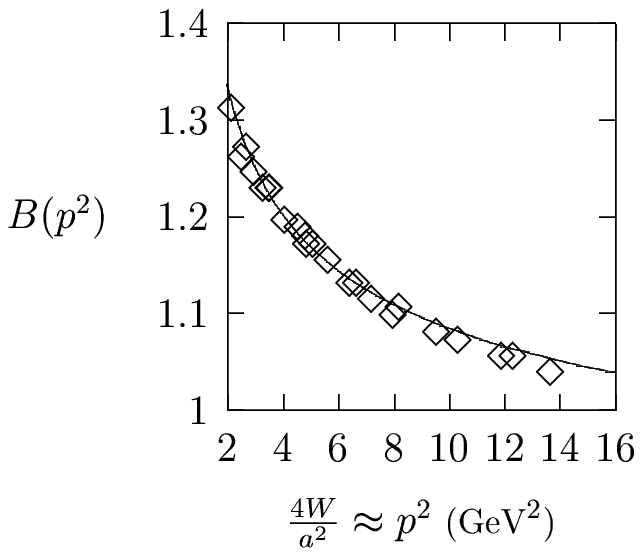,height=2.0in}\hglue 0.5cm
\psfig{file=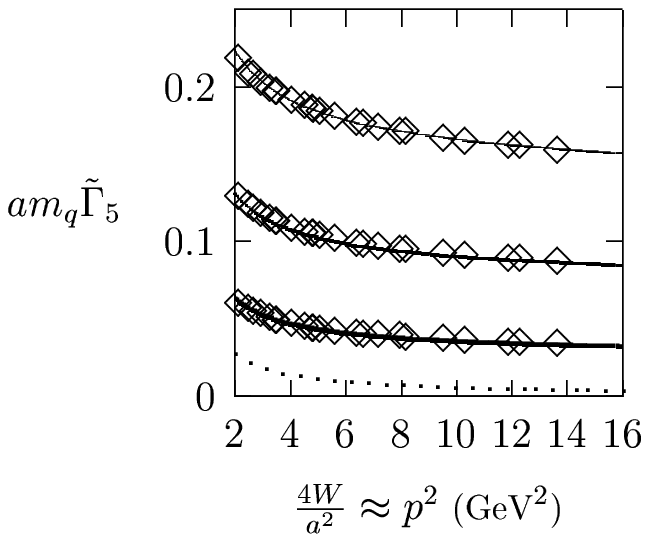,height=2.0in}
}
\caption{
(a) The value of $B(p^2)$ in
eq.~(\ref{eq:fitform})
from our extrapolation
at $\kappa_c$, compared with eq.~(\ref{Bp2});
(b) Our fit to the lattice data in the region of high
$p^2$ for $B_0=1.735$. The plain curves correspond from top to bottom
respectively to $am_q=0.148,~0.078$, and $0.028$, and the dashed one to
the Goldstone boson contribution.
}
\end{figure}

As
$
\Gamma_5 (q=0,p^2)\propto \left[A(p^2)/am_q+B(p^2)\right]
$,
it seems reasonable to
identify
the $A(p^2)$ term as the
Goldstone contribution ({\it i.e.} as the pole in
$am_q$), and the second one as
the perturbative contribution,
if we are sufficiently close to the chiral limit.
The continuum Ward identity enables us to borrow the two-loop
renormalisation group improved factors, from those calculated in the
case of the propagator by the Rome group \cite{allton}
for the perturbative part:
 \beq
B(p^2)=B_0 \times [\alpha_s(p^2)]^{4/11} \left[1+8.1\ {\alpha_s(p^2)\over 4\pi}
\right]
\label{Bp2}
\eeq
and by Pascual and de Rafael \cite{pascual} for the non-perturbative part:
\beq
A(p^2)=A_0 \times {[\alpha_s(p^2)]^{7/11}\over a^2p^2} \left[1+22.0\
{\alpha_s(p^2)\over 4\pi}\right]
\label{Ap2}
\eeq
with $A_0$ and $B_0$ some constants.
These two-loop corrections are valid for the MOM renormalisation
scheme, in the Landau gauge, but with $\alpha_s$ taken as the
$\overline{\small\rm MS}$
coupling constant\footnote{for which we use the expression $4\pi/\alpha_s(q^2)=
11\log\left({q^2\over \Lambda^2_{\small\overline{\small MS}}}\right)
+{102\over 11}\log\left(\log\left({q^2\over
\Lambda^2_{\small\overline{\small MS}}}\right)\right)$.}.
As already mentioned, we take
$a\Lambda_{\small\overline{\small MS}}=0.174$ at $N_F=0$ from the three gluon
coupling
measurement with asymmetric momenta
\cite{pittori1,quadrics}.
Note that, a priori, the evolution formulae properly apply
to the renormalised propagator, which is the bare one divided
by $Z_\psi(p^2)$ \footnote{In this letter, we use the
convention of ref.\cite{rome}
for the $Z$'s.}; but, at least theoretically
and in the continuum, $Z_\psi(p^2)$ should evolve very slowly, since
$\gamma_\psi=0$ at one loop in the Landau gauge, and the two-loop
correction seems also to have very little effect \cite{franco}.
Lattice data \cite{martinelli, skullerud} confirm this perturbative argument.
Then, we
conclude that the two-loop corrections should be obeyed by the bare PS vertex
with good
accuracy.

The lattice data turn out to be quite close to the
continuum theoretical expectations near the chiral limit, confirming the above
interpretation of $A$ and $B$.
Indeed one finds
the following:\\
$\bullet$ $A(p^2)$ is behaving remarkably close to $1/p^2$ over a large
interval
of $p^2$, see Fig.~1(a).
{}From the numerical analysis,
the Goldstone contribution appears to be very large. Indeed,
\beq
a^2p^2 A(p^2)
\simeq 0.015\label{p2A}
\eeq
from the lowest point $a^2p^2=0.16$.

On the other hand, we do not see the log factors expected from
the perturbative calculation, eq.(\ref{Ap2}), which remains to be
understood; we shall discuss further problems 
in the perturbative evaluation of the Wilson coefficient
at the end of the paper. \\
$\bullet$ $B(p^2)$ is found to evolve closely to
$\ln (p^2/\Lambda_{QCD}^2)^{-4/11}$,
more precisely it is evolving in good conformity with the
two-loop MOM renormalisation formula quoted above.
We obtain
\beq
B_0=1.735
\eeq
which provides a very good fit to the data for $p^2$ larger than $2$ GeV$^2$,
see
Fig.~2(b).

The Goldstone contribution is felt already at
rather large quark masses and, for physical $u,d$ quarks it is in fact very
large: $A(p^2)$ is larger than the perturbative part $am_q B(p^2)$
even at rather large $p^2$, as shown in Fig.~3(a) and further discussed in
section 4.

\section{Consequences on the short-distance mass from
the lattice
} \label{sec:sdmass}
Let us then recall that the pseudoscalar vertex is an important
ingredient in the method first developed by the Rome group \cite{allton}
to determine the short distance quark masses. One starts from the ``axial"
MOM renormalised quark mass given by:
\beq
        am_{AWI}^{Landau}(\mu^2)=\rho {Z_A\over Z_P(\mu^2)}
\eeq
where $\rho$ is a dimensionless parameter
determined from a ratio of matrix elements
involving the pion and the bare axial and pseudoscalar bilinear operators.
$Z_A$ is the standard renormalisation of the axial current, determined from
the axial Ward identity or approximated through the MOM renormalisation
non-pertur\-ba\-ti\-ve method, or from one-loop perturbation theory. Finally
$Z_P(\mu^2)$ is defined through the MOM renormalisation condition for the
vertex and is determined by the same non-perturbative method from the PS
quark vertex at $q=0$, or again approximated from one-loop perturbation
theory\footnote{Of course, using the PS quark vertex for normalisation is a
particular
way to define the MOM scheme, which, in the non-perturbative regime,
is not necessarily compatible with {\it e.g.} the use of the scalar vertex. See
remarks below.}. Then, from $m_{AWI}^{Landau}(\mu^2)$, one can deduce the
standard short distance masses, for instance the $\overline{\small\rm MS}$
mass, but this
requires to work in the perturbative, short distance, region.\par

Then our findings for the PS vertex have important consequences.
Let us note that, in principle,
the fact that at $q=0$, the PS vertex not only
is influenced by a large Goldstone contribution
(as noted in \cite{rome}, \cite{olrich}), but
really explodes in the chiral limit, is crucial for
the MOM procedure of renormalisation,
since the renormalisation constant $Z_P$ is defined as:
\beq
Z_P^{MOM}(\mu^2)= {Z_{\psi}(\mu^2)\over  \Gamma_5(q=0,p^2=\mu^2)}
\eeq
\begin{figure}[t]
\vglue 3.4cm
\hglue 6.5cm (a)\hglue 6.0cm (b)\\
\vglue -2.9cm
\centerline{
\psfig{file=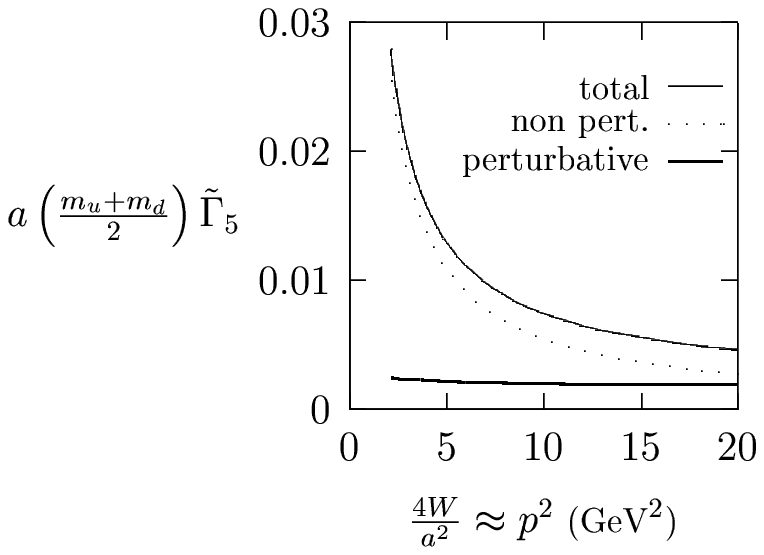,height=2.0in}\hglue 0.5cm
\psfig{file=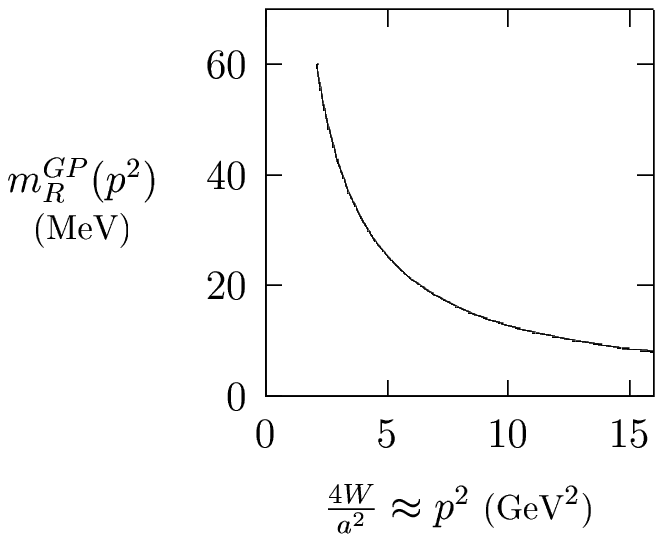,height=2.0in}
}
\caption{ (a) The value of $am_q \tilde\Gamma_5(q=0,p^2)$ for light quarks;
(b) The value of the dynamical $u, d$ masses.
}
\end{figure}
This definition ensures that
\beq\Gamma_5^R(q=0,p^2)=1 {\ \rm at}\ p^2=\mu^2\label{normag}\eeq
 and one concludes from it
that
$Z_P$ has a trivial chiral limit at fixed $p^2$.
Indeed, near $\kappa_c$, since
$ m_{\pi}^2 \propto m_q$ and $
\Gamma_5(q=0,p^2) \propto 1/m_q$,
$Z_P^{MOM}$ tends to zero, or its inverse goes
to infinity.

\subsection{Calculation of $Z_P$}
One can translate the above fit
of $\Gamma_5$  into an expression for $Z_P^{MOM}$,
or rather its inverse which is more directly
physical:
\beq
{1\over Z_P^{MOM}(p^2)}={\Gamma_5(p^2)\over Z_\psi(p^2)}
=
{A_Z(p^2)\over am_q} +B_Z(p^2)
\label{ZPMOM}\eeq
where $A_Z(p^2)=A(p^2)/[CZ_{\psi}(p^2)]$
and $B_Z(p^2)=B(p^2)/[CZ_{\psi}(p^2)]$.

This is to be contrasted with usual fits, which
assume that $Z_P$ is linear in $am_q$ of eq.~(\ref{amq})\footnote{D. Becirevic
\cite{damir} has now done
a fit along the above lines with the data of the Orsay-Rome group and found
roughly similar conclusions.}.
One notes that since $Z_{\psi}(p^2)$ is weakly dependent on $p^2$
and on $\kappa$, and close to 1, the expression is
quite similar to the preceding one: it consists in a first
term which is approximately in $1/m_q$, and the second one
which is approximately
constant\footnote{One could be worried
by the fact that the
dependence of $Z_{\psi}$ on $\kappa$ could generate from
$A_Z/[(1/\kappa-1/\kappa_c)/2]$ an additional contribution to $Z_P^{-1}$
independent of
$\kappa$. However, it can be seen that since $A_Z$ decreases rapidly, this
contribution is not large with respect to the one coming from $B_Z$.}.
Just as for the $\Gamma_5$ vertex,
the former corresponds to the non perturbative
Goldstone contribution while the latter, $B_{Z}$, corresponds to the short
distance contribution, as measured by the lattice numerical simulation,
including all the orders of perturbation theory by a non perturbative
method.
To give numbers, we need now values for $Z_{\psi}$, which we borrow from
the data of the Rome group \cite{martinelli}, with an improvement
procedure trying to parallel as much as possible the one followed by
Rakow for the scalar part.
However, we would like to emphasize that all the qualitative conclusions of
this
paper are independent of precise values of $Z_{\psi}$ as long as they stay
around $1$.

At $a^2p^2=1$ which is close  to the standard reference
point $p=2~GeV$, and at $\kappa=0.1342$, which is the $\kappa$ closest  to the
chiral limit, we estimate:
\beq
Z_{\psi} \simeq 0.85
\eeq
One has then numerically:
\beq
A_Z \simeq 0.023,~~~~~~~~~~~~~~ B_Z \simeq 1.88
\eeq
We emphasize that this
identification of the two contributions does not rely on
Boosted Perturbation Theory (BPT), but purely on numerical
simulations.
We shall refer to $B_{Z}$ as the ``short distance contribution''
not to be confused with its one-loop 
perturbative estimate.
Now, we
shall compare to BPT estimates here and
in the following for illustrative purposes only.
$B_Z$ is indeed very close to the one-loop
standard BPT evaluation
$Z_P^{-1}(a^2p^2=1)=1.7$ (from $Z_P=0.59$) in the chiral
limit, with $g^2=1.6822$, whereas
the non-perturbative term contributes
around $0.023/am_q \sim 0.8$ on a total of 2.7.
Therefore, the departure
of $Z_P$ from its one-loop perturbative evaluation seems
to be essentially due to the Goldstone boson contribution.
Higher-order radiative corrections  do not seem
to be very large\footnote{These two possible explanations were
suggested in ref.~\cite{rome}.}. However, these conclusions
could be sensitive to details of the data or of the fits.

\subsection{ Calculation of  the $\overline{\small MS}$ mass}
We can now convert our results for the renormalised mass
$m_{AWI}^{Landau}$
into a calculation of the $\overline{\small\rm MS}$
mass.
In ref.~\cite{gimenez}, it is suggested that the use of the MOM
non-perturbative
determination of $Z_P$ (with linear extrapolation in $\kappa$)
in the AWI method improves the results for the $\overline{\small\rm MS}$ mass
with
respect to
previous determinations of $Z_P$ by one-loop perturbative calculations.

However, in principle, to make the conversion to a short distance mass,
we must still make sure that we work in the perturbative regime.
Now, we have isolated the
Goldstone boson contribution which
is {\it essentially} non-perturbative as it does {\it not}
correspond to higher order contributions but rather
to power corrections.
The fact that the non-perturbative estimate
of the full $Z_P$ differs sizeably from the short distance
$B_{Z}$ already at the measured kappas,
is a signal that
it is not presently possible to work at $p^2$ high enough for the
Goldstone contribution to be negligible. Hence, we must first
subtract it from $Z_P^{-1}$.
\beq
\left[Z_P^{Subtr}(p^2)\right]^{-1}=\left[Z_P^{-1}(p^2)\right]-{A_Z(p^2)\over
am_q}=B_Z(p^2)
\eeq
Numerically, the remaining short distance $B_{Z}$
is in fact close
to the one-loop BPT
estimate of $Z_P$, as we have just seen.
Indeed, at $a^2p^2=1$,
with this subtraction and using again $Z_{\psi} = 0.85$
near the chiral limit, we find $Z_P^{Subtr}=1/1.88=0.53$
which corresponds to the fully resummed short-distance
contri\-bu\-tion determined directly from the lattice data\footnote{
We take $Z_P^{Subtr}$  approximatively
independent of mass, except for a small variation of  $Z_{\psi}$.
One would otherwise require a fit of $m_q \Gamma_5$ with one more term
in $m_q$.}.
To get an estimate of the consequences on the light quark masses,
we use \cite{gockeler} (which uses
the notation $a \tilde m$ for $\rho$) where 
one finds
$\rho \sim am_q$,
with $am_{u,d}=0.001836$, and we take
$Z_A=0.79$ \cite{luscher}. We then find:
\beq
         a~m_{u,d}^{Landau} \sim 0.0027
\eeq
Converted into the $\overline{\small\rm MS}$
scheme through \cite{allton}
\beq
m_q^{\overline{MS}}=m_q^{Landau}~\left(1-{16 \over 3} {\alpha_s \over
4\pi}\right)
\eeq
with $\alpha_s(a^2p^2=1)=0.25$ at two loops, this gives:
\beq
	a~m_{u,d}^{\overline{MS}} \sim 0.0024
\eeq
therefore about $4.6$ MeV at $N_F=0$.
One would obtain about $6$ MeV
if one used the full MOM non-perturbative estimate of $Z_P$ linearly
extrapolated to the chiral limit
and $4.2$ MeV from one-loop BPT. Note that these
numbers are only indicative; in view of the many uncertainties
in the subtraction procedure, we do not try to discuss the
other sources of error necessary to give a real determination
of the mass. Our aim is only to underline the necessity of the subtraction
of the Goldstone contribution.

Despite the fact that in this case the two methods lead to comparable results,
the subtraction method just described, which does not rely on perturbation
theory but which rather uses the lattice data directly,
is in general superior to that based on
the BPT estimate, and
knowledge of the full $Z_P^{MOM}$ is in general necessary
even if we aim at measuring short distance quantities.

Indeed, the BPT method has the following drawbacks.
Firstly, the unknown higher-order perturbative corrections
may be large. Secondly, the one-loop
perturbative evaluation can be tadpole-improved
in many ways, potentially
leading to very different estimates. Finally,
the one-loop estimate, even if tadpole-improved,
does not automatically follow the behavior dictated by the
renormalisation group; the problem is then cured
by taking the one-loop estimate at some momentum,
and then imposing the renormalisation group
evolution for the other momenta, but this is
obviously presenting a rather arbitrary
choice of a privileged point.

On the other hand, the subtraction method used here
avoids these problems.
It amounts in general to a non-perturbative measurement
of
the $Z$'s, followed by the evaluation and removal of
the pole contributions.
As shown above in Fig.~2(b), this
procedure leads to a result which evolves
per se according to the renormalisation group.

The evaluation of the pole contribution from the pion
is especially easy, because
of its particular singular nature at $m_q=0$.
Other pole contributions  are expected
to be smaller, because they are regular in
the chiral limit. If one were to subtract them,
one could only rely on the
expected power behavior of the particle vertex
function, and their extraction
would thus be more difficult.

\section{Consequences for the renormalised mass \`a la Georgi-Politzer}
\label{sec:propagator}
In this section, we show that the Goldstone boson contribution to the PS
vertex, which is only parasitical in the calculation of $\overline{\small\rm
MS}$ masses,
and
has to be subtracted as shown in the previous sections, retains an important
physical meaning, as can be seen through the use of other definitions of
renormalised quark masses.

\subsection{Physical relevance of the full renormalised axial mass}
One should remember that the full axial
MOM renormalised mass, which is calculated through:
\beq
        am_{AWI}^{Landau}(p^2)=\rho Z_A [Z_P(p^2)]^{-1}  \label{masseaxiale}
\eeq
with $Z_P(p^2)$ not submitted to the above
subtraction, retains a physical significance by itself,
since it is the
mass defined
through the divergence of the axial current,
with the corresponding natural renormalisation condition which
consists in setting the pseudoscalar vertex $\Gamma_5^R$
to $1$ on quark states at the renormalisation scale\footnote{Recall then that
the
scalar vertex should be consistently normalised
through the Ward identities, as repeatedly emphasized by
the Rome group, therefrom one sees that $Z_S/Z_P$ must be
independent of $p^2$, in obvious contradiction with imposing
simultaneously a similar MOM condition for the scalar vertex.}.
We stress that in contrast to the
standard $\overline{\small\rm MS}$ current mass, it does not vanish  in the
chiral limit, because the chiral limit $\rho \to 0$ is compensated by
the pole in $Z_P^{-1}$. The meaning  of this last result will  be further
developed in the next subsection.
Of course, it is unpleasant to have renormalised
quantities with a rather queer behavior in the
chiral limit. Indeed, for instance, hadronic matrix elements
of the renormalised pseudoscalar density
which do not have a pion pole should tend to zero
in the chiral limit- and therefore also, the ones of
the scalar density defined in accordance with the
Ward identity. But as we hope to have shown,
this is an unavoidable consequence of the non-perturbative method
as applied to the pseudoscalar density.
Of course, one could avoid this by preferring
the corresponding MOM normalisation condition for
the scalar vertex.
But then, one must remember another aspect
of the axial
MOM renormalised mass, which gives it another important
physical significance, and which we shall now discuss.

\subsection{ Relation with the mass as defined by the propagator}
It can indeed be easily shown
that the ``axial"  renormalised mass is also
essentially identical to a
standard renormalised mass defined through the scalar part of
the propagator, as
becomes obvious through the renormalised
axial Ward identity
at zero transfer (this has been recalled in
the talk of Rakow); in fact it is essentially identical to
the Georgi-Politzer mass. Let us indeed write the
Ward identity\footnote{Here we use  eq.~(\ref{normalisation}) as a constraint
allowing to calculate the r.h.s. from
the l.h.s. We do not require the independent input of the scalar part of the
propagator data. As a side remark, as a check of the validity of
eq.~(\ref{normalisation}) on the
lattice QCDSF propagator data, we note that it would imply for the $C$ defined
above $C=Z_A\rho/am_q$, which is not far from their $C=0.75$.}:
\beq
m_{AWI}^{Landau}(\mu^2) \Gamma_5^R(q=0,p^2,\mu^2)=
{Tr[S_R^{-1}(p,\mu^2)]\over 4}\label{normalisation}
\eeq
Here, it is assumed of course that the propagator is also
calculated in the Landau gauge, and that the renormalisation
of the vertex and of the propagator are both
consistently performed according to the MOM
scheme. Therefore the propagator is normalised in the Euclidean region
according to\footnote{We disregard here the difference between
this standard MOM condition for $Z_{\psi}$ and the derivative MOM condition
derived from the vector Ward identity by the Rome group \cite{rome}, which
seems to
lead numerically to very small differences.}:
\beq
 S_R^{-1}(p,\mu^2)
=i\not p +m_R^{GP}(\mu^2) ~~~~~\mbox{at} ~~~~~p^2=\mu^2
\label{normalisation2}
 \eeq
which is nothing else than the Georgi-Politzer
renormalisation condition.

Setting $p^2=\mu^2$, one obtains from eqs.~(\ref{normag}),
(\ref{normalisation})
and (\ref{normalisation2}):
\beq
m_{AWI}^{Landau}(\mu^2)={Tr[S_R^{-1}(p,\mu^2)]\over 4}\vert_{p^2=\mu^2}=
m_R^{GP}(\mu^2) \label{identity}
\eeq
therefore the announced identity of $m_{AWI}^{Landau}(\mu^2)$ and the
Georgi-Politzer mass function $m_R^{GP}(\mu^2)$ is derived.

Through eq.~(\ref{identity}), the $1/p^2$ power contribution  in $Z_P^{-1}$
corresponding
to the Goldstone  boson is related  to a similar contribution in the scalar
part of the propagator, which represents
a  dynamically generated mass for light quarks, though
off-shell, gauge-dependent  and Euclidean. This is a well-known
\cite{NJL}
signal of the spontaneous breakdown of the chiral symmetry. We can then
translate our knowledge of the PS vertex into information on this dynamical
mass.

\subsection{Physical consequences}
Numerically, for the non-perturbative contribution,
which corresponds to the chiral limit of $m_{AWI}^{Landau}$,
one finds at $a^2p^2=1$,  from eq.~(\ref{masseaxiale}),
with, as before, $Z_P$ from eq.~(\ref{ZPMOM}),
$Z_A=0.79$
and $\rho\sim am_q$ \cite{gockeler},
\beq
a~m_R^{GP}(a^2 p^2=1) \sim 0.018
\eeq
therefore around $34~\Mev$ at $p=1.9~\Gev$.

We have obtained analogous results directly from the scalar part of the bare
improved propagator as given in~\cite{rakow}.
Extended considerations
and estimates on the propagator will be given in a forthcoming
paper \cite{cudell}.

At large $p$, we want to stress that the non-perturbative contribution
 to the mass
remains of the order of the $u,d$ perturbative masses and
even larger, see Fig.~3(a).
It must be emphasized that this result is rather safe, since it does not
depend critically on improvement procedures.

The sign is as expected, but the magnitude is much larger than what one would
expect from the
estimate by the $q \overline{q}$ condensate and a perturbative
calculation of the Wilson coefficient\cite{pascual}:
\beq
m_R^{GP}(p^2)=-{4 \over 3}\pi \alpha_s {<0|\bar{q} q|0>\over p^2}
\eeq
Indeed, with $\alpha_s^{\overline{MS}}(a^2p^2=1)\sim 0.3$ (one-loop),
and with a standard $\overline{\small\rm MS}$ renormalised
evaluation of the $\bar{q} q$ condensate of $-(225~$MeV$)^3$,  at $\mu=1~GeV$,
rescaled at $a^2p^2=1$ by a factor 1.18,
one would find an answer lower
by a factor ten.
However, a large value is inevitable
if we admit as usual that at moderately low $p^2$ the mass
is of the order of a constituent mass, i.e. several
tens of MeV, and if we
take into account that the decrease is only in $1/p^2$,
as found with remarkable accuracy on the lattice data
of the QCDSF group.

That a value of several tens of MeV is needed at low $p^2$ is
confirmed by direct lattice calculations of the propagator,
to be compared with our estimate of Fig.~3(b):\\
$\bullet$ in configuration space, the propagator in time $S(t)$ has been found
exponential over a large range of $t$, with a coefficient of the exponential
around $300~$MeV \cite{skullerud,bernard}; \\
$\bullet$ in momentum space,
at the lowest points, corresponding to $p^2\rightarrow 0$,
where it should be insensitive to the improvement,
the scalar part of the inverse propagator is found to be around similar
values of $200-400~$MeV \cite{rakow,martinelli}, for the lowest mass
$\kappa=0.1342$.
A large value for the coefficient of $1/p^2$ was also found in a phenomenology
of the pion as a Goldstone boson \cite{stokar}, based on an assumption:
\beq
                   m_R^{GP}(p^2)={4(m_D)^3\over p^2}
\eeq
where $m_D$ is a free parameter. The phenomenology seems to require 
$m_D \approx 300~$MeV, $m_R^{GP}(p^2) \sim 27~$MeV at $p=2~$GeV, in agreement
with the above estimate.

It must be emphasized that this large non-perturbative
contribution does not contradict directly the sum rule calculation
of correlators, which uses a perturbative evaluation of the
quark propagator. Indeed, as in the calculation
of propagator by Politzer,
in sum rule calculations non-perturbative contributions
are consistently added through condensates. Moreover,
as emphasized by 
Pascual and de Rafael \cite{pascual}, the condensate
contributions are quite different for the quark propagator
and the correlators, therefore our finding for the full
quark propagator does not have a direct impact on sum rules.

Of course, the discrepancy with the na\"\i ve perturbative
estimate  of the  Wilson coefficient
is nevertheless worrying and deserves further reflexion
since the latter is known to work in general.
One could imagine that the perturbative
calculation of the Wilson coefficient is not valid for some
particular reason.
In this direction, one must observe that the two-loop
correction is very large, around $50 \%$ of the first order at $a^2p^2=1$.
It may be therefore that the perturbative expansion happens to fail.

\nonumsection{Conclusion}
The lattice numerical calculations can be seen
as the triumph of the general theoretical predictions
of the 70's for the quark pseudoscalar vertex.
Nevertheless, a striking and unexpected
feature of the lattice data is the
very large size of the Goldstone boson contribution to the pseudoscalar
vertex. This corresponds, through the Ward identity, to a
very large non-perturbative contribution to the
renormalised mass function of Georgi and Politzer, by far larger than
what is expected from the quark condensate and a perturbative
evaluation of the Wilson coefficient, but in agreement
with other physical expectations.

These large non-perturbative contributions
then give a warning that there
may be a possible problem with the use of lowest
order perturbation theory in
the estimate of the Wilson coefficient of condensates.
This question requires more investigation.

The Goldstone contribution must be subtracted from
the pseudoscalar vertex to calculate the short distance mass
from a normalisation of this pseudoscalar vertex. This has important
numerical consequences.
The short distance $Z_P$ to be used
should be sizeably larger
than the one measured directly on the lattice.

Finally, it must not be forgotten that
lattice artefacts may still be large
at $\beta=6$, , and that the vertex has not
been improved with rotations
as an off-shell Green function,
therefore one can expect large uncertainties
on the quantitative estimate of the effect at large $p^2$.

\noindent {\it Note  added  in proof: When  completing this paper, we became
aware of
the work of the JLQCD collaboration, presented at Denver conference,
\cite{JLQCD} where certain parallel conclusions on $Z_P$ have been drawn from
lattice staggered fermion data. Paul Rakow has also drawn our attention to
ref.~\cite{rakow2}
, where connected observations on PS vertex and the scheme of Pagels are
made.}

\noindent{\bf Acknowledgements:}  We are very grateful to Paul Rakow  for
providing us with the data on the pseudoscalar  vertex, as well  as for very
useful related information and
discussions, to Guido Martinelli for invaluable encouragement,
in particular for many discussions, precious information and  ideas  or notes
on the
current work  of the Rome group on quark masses and on the quark propagator,
and
for providing us with their data on the propagator.
A. L. would also like
to thank Damir Becirevic with whom much of this paper has been intensively
discussed, for many informations, and finally all the lattice group, and the
quark model group at Orsay for their constant and friendly help and
discussions, as well as J. Skullerud. C. P. acknowledges illuminating
discussions with  G.C. Rossi, warmly thanks J. Cugnon and
the ``Groupe de physique nucl\'eaire th\'eorique de
l'Universit\'e de Li\`ege'' for
kind hospitality and acknowledges the partial support of IISN.

\vbox{\nonumsection{References}
\vglue -16pt}


\begin{thebibliography}{1}
\bibitem{allton} C. Allton et al., Rome group, {\it Nucl. Phys.} {\bf B431}
(1994) 667.
\bibitem{rome} G. Martinelli et al., Rome group, {\it Nucl. Phys.} {\bf B445}
(1995) 81, e-print hep-lat/9411010.
\bibitem{alpha} ALPHA collaboration, S. Capitani et al.,
hep-lat/9709125, talk given
by M. L\"uscher at the International Symposium on
Lattice Field Theory, July 22-26, 1997, Edinburgh.
\bibitem{lane} K. D. Lane, {\it Phys. Rev.} {\bf D 10} (1974) 2605.
\bibitem{pagels} H. Pagels, {\it Phys. Rev.} {\bf D 19} (1979) 3080.
\bibitem{politzer} H. D. Politzer, {\it Nucl. Phys.} {\bf B117} (1976) 397.
\bibitem{vladikas} G. Martinelli, S. Petrarca, C.T. Sachrajda, and A. Vladikas,
{\it Phys. Lett.} {\bf B311} (1993) 241, Erratum: {\it ibid} {\bf B317} (1993)
660.
\bibitem{cudell} J.R. Cudell, A. Le Yaouanc and C. Pittori, ``Lattice Quark
Propagator and Quark Masses: an Extensive Analysis
 of  Available Data'', article in preparation.
\bibitem{pascual}  P. Pascual and E. de Rafael, {\it Zeit. Phys.}
{\bf C 12} (1982) 127.
\bibitem{rakow} Paul Rakow, for S. Capitani et al., QCDSF collaboration,
{\it Nucl. Phys. B} (Proc. Suppl.) {\bf 63} (1998) 871, e-print
hep-lat/9710034.
\bibitem{SW} B. Sheikholeslami and R. Wohlert, {\it Nucl. Phys.} {\bf B259}
(1985) 572.
\bibitem{BS} G.S. Bali and K. Schilling, {\it Phys. Rev.} {\bf D47} (1993) 661.
\bibitem{pittori1} D. Henty et al., in the proceedings of the 1995 EPS HEP
conference (Brussels, 1995), p. 239, e-print hep-lat/9510045.
\bibitem{quadrics} J. P. Leroy, Orsay Quadrics group, private communication.
New results are given in hep-ph/9810322.
\bibitem{franco} E. Franco and V. Lubicz, preprint ROME-I-1198-98 (March 1998),
e-print hep-ph/9803491.
\bibitem{martinelli} G. Martinelli, propagator data of the Rome group,
private communication.
\bibitem{skullerud} J. Skullerud, for the UKQCD collaboration, {\it Nucl. Phys.
B} (Proc. Suppl.) {\bf 42} (1995) 364, e-print hep-lat/9412014.
\bibitem{olrich}  H. \"Olrich, for M. G\"ockeler et al., QCDSF collaboration,
{\it Nucl. Phys. B} (Proc. Suppl.) {\bf 63} (1998) 868, e-print
hep-lat/9710052.
\bibitem{damir} D. Becirevic, private communication.
\bibitem{gimenez} V. Gim\'enez et al., hep-lat/9801028v2.
\bibitem{Heatlie} G. Heatlie et al., {\it Nucl. Phys.} {\bf B352} (1991) 266.
\bibitem{gockeler} M. G\"ockeler et al., QCDSF collaboration, {\it Phys. Rev.}
{\bf D57} (1998) 5562.
\bibitem{luscher} M.  L\"uscher, S. Sint, R. Sommer, and H. Wittig, {\it Nucl.
Phys.} {\bf B491} (1997) 344, e-Print Archive: hep-lat/9611015.
\bibitem{NJL} Y. Nambu and G. Jona-Lasinio, {\it Phys. Rev.} {\bf 122} (1961)
345 and {\bf 124} (1961) 246.
\bibitem{georgi} H. Georgi and H.D. Politzer, Phys. Rev. {D 14} (1976) 1829.
\bibitem{bernard} C. Bernard, D. Murphy, A. Soni and K. Yee, {\it Nucl. Phys.
B} (Proc. Suppl.) {\bf 17} (1990) 593.
\bibitem{stokar} H. Pagels and S. Stokar, {\it Phys. Rev.} {\bf D 20} (1979)
2947.
\bibitem{JLQCD} N. Ishizuka, for the JLQCD collaboration, e-print
hep-lat/9809124.
\bibitem{rakow2} M. G\"ockeler et al., preprint DESY-98-097, July 1998
and e-print hep-lat/9807044.
\end{thebibliography}
\end{document}